\documentclass[conference]{IEEEtran}
\usepackage{cite}

\ifCLASSINFOpdf
  \usepackage[pdftex]{graphicx}
  \graphicspath{{./figures/}}
\else
   \usepackage[dvips]{graphicx}
   \graphicspath{{./eps/}}
\fi
\usepackage{enumerate}

\usepackage{array}

\usepackage{tabularx}

\ifCLASSOPTIONcompsoc
 \usepackage[caption=false,font=normalsize,labelfont=sf,textfont=sf]{subfig}
\else
  \usepackage[caption=false,font=footnotesize]{subfig}
\fi
\usepackage{url}

\begin{document}
%
\title{Simulation Study on Collaborative Content Distribution in Delay Tolerant Vehicular Networks\vspace{-1.5ex}}

\author{\IEEEauthorblockN{Rusheng Zhang \IEEEauthorrefmark{1}, Bo Yu \IEEEauthorrefmark{2}, and Hariharan Krishnan \IEEEauthorrefmark{2}}
\IEEEauthorblockA{\IEEEauthorrefmark{1} Department of Electrical and Computer Engineering,
Carnegie Mellon University,
 Pittsburgh, PA 15213-3890, USA}
 \IEEEauthorblockA{\IEEEauthorrefmark{2} General Motors Research \& Development, Warren, Michigan 48092, USA}
 \IEEEauthorblockA{Emails: rushengz@andrew.cmu.edu, \{bo.3.yu, hariharan.krishnan\}@gm.com}
}


%


\maketitle

\vspace{-0.5in}

\begin{abstract}
Modern vehicles are equipped with more and more sophisticated computer modules, which need to periodically download files from the cloud, such as security certificates, digital maps, system firmwares, etc. Collaborative content distribution utilizes V2V communication to distribute large files across the vehicular networks. It has the potential to significantly reduce the cost of cellular-based communication such as 4G LTE. In this report, we have conducted a simulation study to verify the feasibility of a hybrid cellular and V2V collaborative content distribution network.
In our simulation, a small portion of the simulated vehicles download the file directly from the cloud via cellular communication, while other vehicles receive the file via collaborative V2V communications. Our simulation results show that, with only 1\% of vehicles enabled with cellular communication, it takes less than 24 hours to distribute a file to 90\% of the vehicles in a metropolitan area, and around 48 to 72 hours to distribute to 99\%. The results are very promising for many delay-tolerant content distribution applications in vehicular networks.
\end{abstract}


%
\IEEEpeerreviewmaketitle

\vspace{-0.2in}
\section{Introduction}
As Dedicated Short-Range Communication (DSRC) radios, vehicle infotainment systems, as well as other computer systems are equipped onto modern vehicles, there is a need to distribute large files from cloud to remote vehicles periodically. For example, these files may include Certificate Revocation List (CRL) files for DSRC radio, firmware, apps and HD maps for infotainment systems, etc. The file size may range from a few kilobytes to hundreds of megabytes. Since most of modern vehicles are equipped with cellular communication devices, such as OnStar LTE, a straight-forward way to distribute these files is to distribute them through the cellular network. However, the cost of cellular communication could be high, due to the large number of vehicles.
	
\begin{figure}[ht]
\centering
\includegraphics[width=3in]{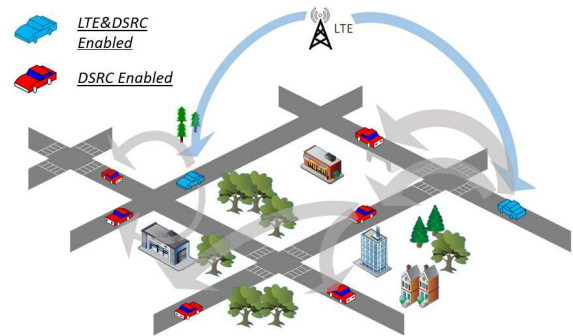}
\caption{A concept figure of collaborative content distribution}
\label{fig_concept}
\end{figure}

An alternative approach to cellular-based file distribution is Collaborative Content Distribution \cite{nandan2005co, lee2006code,johnson2006collaborative,zhang2010roadcast,sathiamoorthy2014distributed,laberteaux2008security}. As opposed to the traditional scenarios described in these papers, which use roadside units as gateways to distribute contents. We are interested in using LTE to distribute contents directly to the vehicles, and further disseminate data through Vehicle-to-Vehicle (V2V) communication. Figure \ref{fig_concept} shows the concept of our proposed scheme in Vehicular Networks. With Collaborative Content Distribution, we selectively enable a small portion of vehicles which use LTE communication to download the files directly from the cloud, and then use those vehicle as seed vehicles. As the seed vehicles move around, they can distribute the files to other non-seed vehicles by V2V communication, such as DSRC. These non-seed vehicles can then become seed vehicles and continue to distribute the files across the vehicular network. Since only a small portion of vehicles use LTE to download the files and most vehicles receive the files via V2V communication, the system will minimize the expensive LTE bandwidth, as well as the cost of cloud server workload. Therefore, service cost can be significantly reduced. 
\vspace{-0.1in}


Files are encoded into small chunks on the cloud with fountain codes. Once a vehicle has collected enough number of chunks, it is able to reconstruct the original file. Notice that though it's possible for each vehicles to encode more file chunks, for security reasons, encoded chunks are distributed only from the cloud with a digital signature, vehicles can retransmit these chunks, but can't generate new chunks. 

In this work, we have conducted a simulation-based study to verify the feasibility of a hybrid collaborative content distribution network and understand the performance changes under different system parameters. The purpose of this simulation study is two-fold as follows:
\begin{enumerate}
\item To estimate a rough value or  bounds for the propagation time in the early phase of the technology deployment; we focus on CRL file distribution, but the results are also referable for other file distribution system.
\item To better understand the sensitivity of different parameters and movement patterns.
\end{enumerate}

\vspace{-0.1in}
\section{Related Work}
Nadon et.al. \cite{nandan2005co} first proposed cooperative downloading in VANETS. In which, they proposed an vehicular network architecture composed of two communications: Vehicle-vehicle and Vechicle-gateway. They proposed SPAWN protocol, instead of traditional swarming protocol such as BitTorrent and Slurpie to adapt to the vehicular environment. Based on which, CodeTorrent\cite{lee2006code} and VENETCODE\cite{ahmed2006vanetcode} are proposed,  which are single-hop P2P file sharing system based on Network coding. Li et.al. \cite{li2011codeon} proposed CodeOn system, improving the performance by using Symbol Level Network Coding (as opposed to Packet Level Network Coding). Shen et.al. \cite{shen2014data} proposed a scheduling based protocol (as opposed to random access), they also utilize Space-Time Network Coding instead of Random Linear Network Coding to improve successful decoding rate. Zhang et.al. designed a popularity aware system called Roadcast improves the content selection mechanism \cite{zhang2010roadcast} by having vehicles not only share the information they need, but also the information popular throughout the network, hence, increasing the overall satisfaction.

Meanwhile, several analysis based on theory and simulations are carried out to evaluate the collaborative downloading systems. Johnson et.al. \cite{johnson2006collaborative} investigated two transmission schemes based on NC, the Synchronous scheme (spatial) and Asynchronous scheme from theoretical analysis and simulation respectively. Firooz et.al. gives a thorough analysis of such Network coding based wireless system \cite{firooz2012collaborative,firooz2013wireless}.  Viriyasitavat et.al. \cite{viriyasitavat2011dynamics} research on the mobility of the vehicles and give connectivity performance over Manhattan Grid (MG) based on simulation, together with an analysis framework. Laberteaux et.al. \cite{laberteaux2008security} investigate the capability of collaborative downloading system for sharing Security Certificate Revocation (CSL). 

The main contribution of the paper is as following:
\begin{enumerate}
\item As opposed to aforementioned systems, which are V2V systems with roadside gateways. Here, we propose a LTE-DSRC hybrid system without stationary roadside gateways (Figure \ref{fig_concept}). 
\item We have conducted a city level simulation, where the map, the number of vehicles, mobility pattern are comparable with reality, thus, giving a credible reference of the performance for future real-world deployment.
\item We also performed simulations over Manhattan Grid (MD) to perform controlled experiments and analyzed the effect of different mobility patterns.
\end{enumerate}

\vspace{-0.1in}
\section{Simulation}
Simulations are performed using SUMO, a microscopic traffic simulation tool \cite{SUMO2012}. A python script interacts with SUMO through the Traci interface and handles the communications between vehicles. Since this is a macroscopic simulation, some details such as control packets are abstracted, we simply consider a uniform file transfer rate. 

Based on different purposes, as shown in Figure \ref{fig_simulation_map}, different maps are used in the simulation:

\begin{figure}[!ht]
\centering
\subfloat[OSM Map used in the simulation of downtown Detroit and Warren, Michigan, USA]{\includegraphics[width=3in]{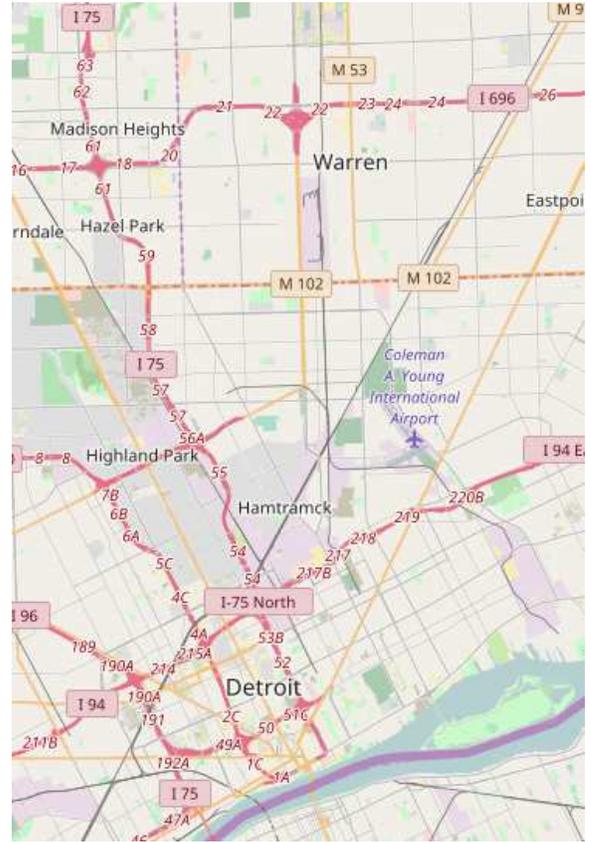}%
\label{fig_map}}
\hfil
\subfloat[An illustration of Manhattan Grids]{\includegraphics[width=2.5in]{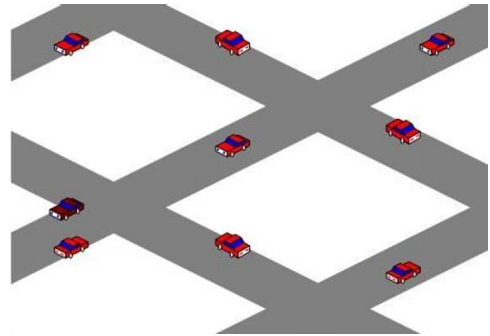}%
\label{fig_MG}}
\caption{Maps used for simulations}
\label{fig_simulation_map}
\vspace{-0.2in}
\end{figure}

\textbf{\emph{Simulation over real world map}}: The first set of simulations is performed on a map exported from Open Street Map (OSM), which is an $18km \times 23km$ rectangle area covering downtown Detroit and Warren, Michigan. Figure \ref{fig_map} shows the area used in the simulation. Two sets of simulation experiments are performed on this map:

\begin{itemize}
  \item The first simulation experiment is performed to evaluate the effect of initial seed rate. Five rounds of simulation are performed to test the performance of the system under different initial seed rate, from 0.02\% to 10\%;
  \item The second simulation experiment is performed to evaluate the effect of the number of total cars. Five rounds of simulation with different car numbers, from 3000 cars to 11000 cars within this area, are performed.
\end{itemize}

In our simulation, each car is assigned with random trips at random times, the expectation of number of trips for a car is 3, namely each car will have on average 3 trips a day, resulting a car occupancy rate of 3-5\%, which agrees with related researches \cite{parked}. This is an average number, meaning there will be some cars having fewer than 3 trips or even no trips at all in a day while some other cars having more than 3 trips a day. The maximum trip distance is assigned as 10km.

The communication distance is set to 100m. While the distance between two cars is less than 100m, cars start to communicate with each other. If one car has data chunks the other car doesn't have, a data transmitting link is setup and transmission starts. The file size in the simulation is 400kB, which is the typical size of a CRL file. We assume fountain codes are used to encode files \cite{ahmed2006vanetcode,li2011codeon,firooz2012collaborative,firooz2013wireless,sathiamoorthy2014distributed}. The file is encoded into 450 chunks on the cloud, and a seed vehicle will receive all these 450 encoded chunks from the cloud via 4G LTE connections. Once these file chunks are distributed into the vehicular network, a non-seed vehicle needs to collect any 300 chunks in order to reconstruct the original file. The data rate of the DSRC radio is set to 6Mbps, which is a typical transmitting rate for DSRC radio.


\textbf{\emph{Simulation over Manhattan grid (MG)}}: Figure \ref{fig_MG} shows a visual example of Manhattan grid, where roads are arranged to form rectangles. Though it's ideal to perform all simulations over the real street map, when it comes to researching effect of unevenly distributed traffic demand, a real world map simulation is too complex and too slow to perform. On the other hand, a Manhattan grid is clean and simple for this purpose, and is perfect for conducting controlled experiments. Hence, a simulation over Manhattan grid is performed to test the effect of unevenly distributed traffic. Several simulations are performed, assigning different rate of car (from 0\% to 100\%) preferring to take some certain 'main roads'. Vehicles randomly choose a destination, but some vehicles will randomly choose a route to the destination while other vehicles will choose a nearest main road to go to the destination. By changing the proportion of these 'main-road' vehicles, we will have an idea on how unevenly distributed traffic demand will affect the final results.



\begin{figure}[!ht]
\centering
\subfloat[Effect of initial seed rate when transfer rate is 800kb/s]{\includegraphics[width=3in]{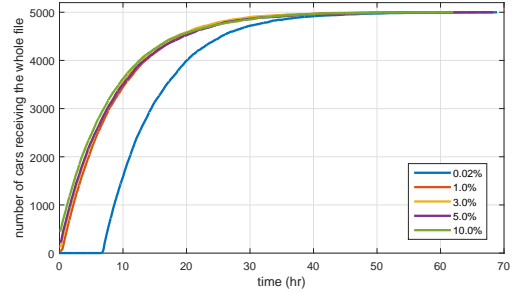}%
\label{fig_600_init}}
\hfil
\subfloat[Zoomed in version of (a) to the first 24 hours from simulation starts]{\includegraphics[width=3in]{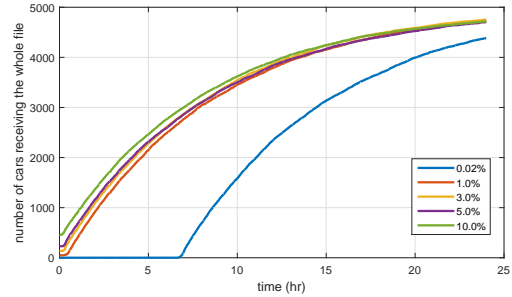}%
\label{fig_600_init_zoom}}
\caption{Effect of initial seed rate.}
\label{fig_600_init_full}
\vspace{-0.2in}
\end{figure}

\vspace{-0.1in}
\section{Results and Discussion}
\subsection{Effect of Initial Seed Rate}

Figure \ref{fig_600_init_full} shows the simulation results for different initial seed rates. While Figure \ref{fig_600_init} shows the results of the whole simulation of 72 hours, the most informative part of the simulation is the first 24 hours and this part is plotted in Figure \ref{fig_600_init_zoom}). The results show that within 12 hours (even shorter when initial seed rate is higher), more than half of the vehicle will complete data transferring (i.e., they receive enough chunks to reconstruct the file). From the figure, we see the trends of all curves follow the same pattern, which is very steep from beginning and becomes flat gradually. This concave shape results in the fact that the initial seed rate only makes visible difference at the beginning of the file file propagation. For example, from the figure, one can clearly observe that though initial seed rates will affect the time of 50\% completion, there is only a slight difference for the time of 80\% completion. At the end of the first day, even with an initial seed rate of as small as 0.02\%, the system is able to have more than 90\% of completed vehicles.  

\begin{table}[h]
\renewcommand{\arraystretch}{1.3}
\caption{Simulation results in different initial seed rate}
\label{table_init}
\centering
\begin{tabularx}{\linewidth}{|X||X|X|X|X|X|}
\hline
Initial Seed & 10\% & 5\% & 3\% & 1\% & 0.02\%\\
\hline
Reach 30\% & 2.5 hours & 3.0 hours & 3.2 hours & 3.8 hours & 6.0 hours\\
\hline
Reach 50\% & 5.5 hours & 6.0 hours & 6.3 hours & 6.9 hours & 9.1 hours \\
\hline
Reach 80\% & 13.6 hours & 14.3 hours & 14.5 hours & 14.8 hours & 17.3 hours \\
\hline
Reach 99.9\% & 72-108 hours & 72-108 hours & 72-108 hours & 72-108 hours & 72-108 hours \\
\hline
\end{tabularx}
\end{table}

Table.\ref{table_init} shows more specific details of this simulation. From the table, one can clearly see that the initial seed rate affects the most at the early stage of the file propagation. For example, to reach 30\% penetration rate, 10\% seed rate will need only 2.5 hours while 1 percent seed rate will need around 4 hours, that's 60\% more time cost. Meanwhile, to reach 80\% penetration rate, very small time difference is observed for different seed rate.

The results we obtained here is both interesting and promising, since the main purpose of such a collaborative file distribution system is to save LTE bandwidth.  As the initial seed rate (LTE enable vehicle proportion) doesn't affect hugely, a very small amount of LTE enable vehicles are required, i.e., 1\%. Namely, such a system can save more than 99\% LTE bandwidth, which is of significant value.

\vspace{-0.05in}
\subsection{Effect of Transfer Rate}
\begin{figure}[ht]
\centering
\includegraphics[width=3in]{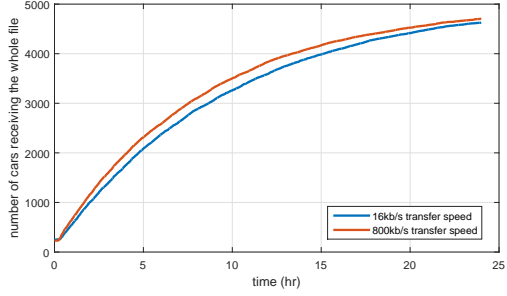}
\caption{Results of difference of transfer rate}
\label{fig_rate}
\vspace{-0.2in}
\end{figure}

The previous results use a typical value of transferring rate of 6MB/s. When the channel condition is severe, the transferring speed of DSRC radio can go as low as 16kb/s. In this section, the results of 16kb/s transferring rate are presented, while other simulation parameters remain the same. The results show a 'lower bound' of the propagation time.
\begin{figure}[!ht]
\centering
\subfloat[Effect of car number displayed in real car number]{\includegraphics[width=3in]{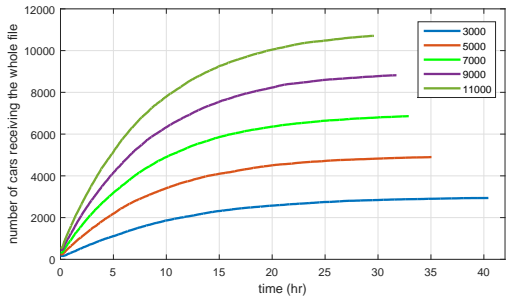}%
\label{fig_cn}}
\hfil
\subfloat[Effect of car number displayed in the completed rate (proportion of cars completed)]{\includegraphics[width=3in]{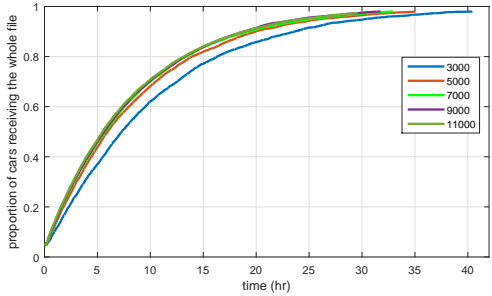}%
\label{fig_cn_pro}}
\caption{Effect of car number.}
\label{fig_cn_full}
\vspace{-0.2in}
\end{figure}

Figure \ref{fig_rate} shows the results of transfer rate of 16 kb/s (blue curve) and 800 kb/s (orange curve) under the initial seed rate of 5\%. The two transfer rates are considered the ideal channel condition and the most severe channel condition respectively, and can be viewed as the 'upper bound' and 'lower bound' of the performance. Observe that the two curves are very close with each other, which indicates that the actual behavior, which should be the curve in between the two curves can be approximately estimated from the two curve. Hence, the simulation results from last section can be considered as an approximation of the system behavior at Detroit area. 

The close behavior of the two cases shows that the CRL file propagation is not sensitive to the transmitting rate. This is because even with only 16kb/s transferring speed, most of the data can already be transferred while two vehicles encounter.

\vspace{-0.05in}
\subsection{Effect of car number}
Figure \ref{fig_cn_full} shows the effect of car number. Figure \ref{fig_cn} shows the results in absolute number of cars completed transferring, Figure \ref{fig_cn_pro} gives the same result in proportion of completed vehicles. From the results we can see that though with more cars, the absolute number of completed vehicles will grow much faster, in proportion, the speed of growth is comparable. This means the propagation speed is proportional to the car number. This is due to the fact that more vehicles in the network will lead to more chances for them to meet each other. In fact, the chance of meeting more vehicles in one trip is exactly proportional to the number of cars in the network.

Even in proportion, with more vehicles, the transfer rate is still higher, this is due to the fact that in the early stage of propagation, if car number is lower, during 1 trip, a vehicle will have less chance to meet vehicles with data it's looking for. So even the chance of meeting other vehicles is proportional, the chance of meeting data-carrying vehicles is smaller. Fortunately, from the simulation, we know this is only a secondary factor and can be overlooked when the car number is from 3000 to 11000. This effect is also smaller when car number increases, this is also understandable since when there are very few cars (approaching 0), it will take an almost infinite time for cars to meet each other, and this factor becomes the critical factor. Meanwhile, when car number increases, this early stage will quickly pass and become a minor factor. This can also be seen from Figure \ref{fig_cn_pro}, that the performance of 3000 car number is the lowest and far from all other results, while 5000, 7000, 9000, 11000 are very comparable.

\vspace{-0.05in}
\subsection{Main Road Effect}
\begin{figure}[ht]
\vspace{-0.1in}
\centering
\includegraphics[width=3in]{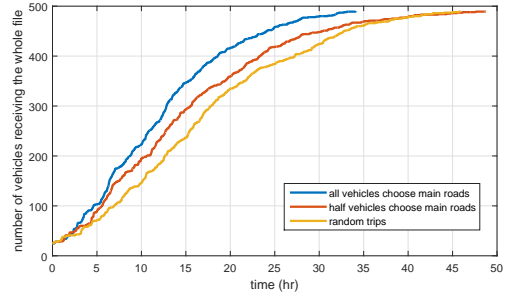}
\caption{The effect of main roads}
\label{fig_mainroad}

\end{figure}

Another simulation performed on a $10\times 10$ Manhattan Grid is performed to see the effect when vehicles choose some preferred 'main roads' instead of randomly choosing routes to destination. In this simulation, 3 vertical roads are assigned to be 'main roads'. All vehicles randomly choose a source and a destination in the map. We consider 3 cases:
\begin{enumerate}
\item All vehicles randomly choose a route from source to destination.
\item Half of vehicles choose the closest main road to travel from source to destination.
\item All vehicles choose the closest main road to travel from source to destination.
\vspace{-0.05in}
\end{enumerate}
Since all other simulation parameters are fixed, the experiment is controlled and we can see exactly how 'main roads' will affect the results.

Figure.\ref{fig_mainroad} shows the effect of main roads, since the simulation is a controlled experiment over an artificial map, absolute value of time and car number is not important. We want to see from this figure the difference in system performance when most of cars are using main roads versus the performance that cars randomly choose roads. In this figure, the green curve shows the performance that all vehicles choose main roads to go to the destination; the orange curve shows the case when half of the vehicles choose main roads; the blue curve shows the case when all vehicles randomly choose a route to the destination. We can see that when more vehicles use main roads, the system performance will become better.

\begin{table}[h]
\vspace{-0.2in}
\renewcommand{\arraystretch}{1.3}
\caption{Simulation results in different movement pattern of vehicles}
\label{table_mainroad}
\centering
\begin{tabularx}{\linewidth}{|X||X|X|X|}
\hline
  & Reach 30\% & Reach 50\% & Reach 80\% \\
\hline
Random Trip & 10.2 hours & 15.4 hours & 27.2 hours\\
\hline
Half cars choose main roads & 7.4 hours & 12.7 hours & 23.3 hours \\
\hline
All cars choose main roads & 6.4 hours & 10.1 hours & 18.6 hours\\
\hline
\end{tabularx}
\vspace{-0.1in}
\end{table}

Table.\ref{table_mainroad} shows more details from the simulation. Though the ‘main roads’ movement pattern also has more effect at the early stage of the propagation, visible difference in time cost can be observed at reaching 80\% penetration rate. This indicates that the ‘main roads’ movement pattern is a more dominating factor than the seed rate and car number. In fact, the time cost for reaching 80\% penetration rate is around 33\% shorter when all cars choose main roads compared with the case when all cars randomly choose a route.

Since the 'main roads' movement pattern is often observed in real world, the results indicate that the collaborative content distribution will be benefited by this factor when deployed in a city of main roads, and the early stage condition will be improved by the movement pattern of traffic.

\section{Conclusion}
\vspace{-0.05in}
From the simulations and results presented in this report, the collaborative content distribution system exhibits great potential in CRL file distribution, with a low initial seed rate (LTE device equip rate). Simulation has shown that within 1 day, 90\% of vehicles will be updated through this system, with only 1\% of equip rate, and the percentage of file completion will reach at least 99\%, if not 100\%, in 2 or 3 days, in a total of 5000 vehicles in Detroit .

The propagation of files in this system will quickly reach a certain percentage, but the last few percentage will take long time to propagate content. This propagation rate is roughly proportion to car number.

In most of cities, the 'main roads' movement pattern will help the content distribution in a major way, at the early and middle stage of the propagation. 

Our simulation results indicate that the Collaborative CRL Distribution system is a promising system that can be deployed in the near future for distributing CRL files as well as lots of other files. The system will save at least 99\% of LTE bandwidth and will finish 90\% of file transfer within 72 hours.







\bibliographystyle{IEEEtran}
\vspace{-0.05in}
\bibliography{IEEEabrv,reference}
%



\end{document}